\documentclass[compress]{elsarticle}

\usepackage{amsmath,amssymb,amsfonts}%

\usepackage{listings}
\usepackage{xcolor}

\usepackage{geometry}
\usepackage{lineno,hyperref}
\usepackage{afterpage}

\modulolinenumbers[1]

\journal{Astronomy and Computing}

\begin{document}

\begin{frontmatter}

\title{A User-Friendly Python Interface for the Numerical Relativity Code AMSS-NCKU \tnoteref{mytitlenote}}


\author[address1,address2]{Chen-Kai Qiao}
\ead{chenkaiqiao@cqut.edu.cn}

\author[address2]{Yi Zheng}
\ead{yizheng0801@163.com}

\author[address3,address4]{Zhou-Jian Cao}
\ead{zjcao@amt.ac.cn}


\address[address1]{School of Physics and New Energy, Chongqing University of Technology, Chongqing, 400054, China}
\address[address2]{School of Mathematical Sciences, Chongqing University of Technology, Chongqing, 400054, China}
\address[address3]{School of Physics and Astronomy, Beijing Normal University, Beijing, 100875, China}
\address[address4]{School of Fundamental Physics and Mathematical Sciences, Hangzhou Institute for Advanced Study (UCAS), \\ Hangzhou, 310024, China}

\begin{abstract}
Numerical relativity has brought about profound and wide-ranging influences on modern astrophysics and gravitational-wave astronomy. In this study, we present a user-friendly Python interface for the numerical relativity code AMSS-NCKU. This interface facilitates the automation of initializing and executing the AMSS-NCKU simulations, as well as the automatic visualization of the output data. The Python interface can significantly reduce the operational complexity of the AMSS-NCKU simulation workflow, lowering the technical barriers for new users. To show the utility of this Python interface, we present two representative examples of numerical relativity simulations (the binary black hole and triple black hole merger processes), obtaining stable numerical results and the expected physical behaviors for black hole systems. 
\end{abstract}


\end{frontmatter}

%\linenumbers

\section{Introduction}\label{sec1}

Over the past several decades of exploration, we have witnessed the observation of the gravitational-wave event GW150914 \cite{GW150914-A,GW150914-B,GW150914-C}, which is a revolutionary achievement in physics and astronomy, bringing us into a new era of exploring the universe through gravitational waves. During more than 10 years of operation, hundreds of gravitational-wave events have been observed by LIGO-Virgo-KAGRA collaborations, whose typical examples include GW170817 \cite{GW170817-A,GW170817-B}, GW190521 \cite{GW190521}, GW230529 \cite{GW230529}, and GW250114 \cite{GW250114-A,GW250114-B}. The accumulation of observational data enables the establishment of gravitational-wave transient catalogs (GWTC) \cite{GWTC-1, GWTC-2, GWTC-2b, GWTC-3, GWTC-4}, providing fundamentally important infrastructures in gravitational-wave astronomy. These catalogs have substantially deepened our understanding on the properties of gravitational waves and the characteristics of extremely compact astrophysical objects.

Numerical relativity has exerted a profound impact on modern astrophysics and gravitational-wave astronomy. During the observational detection of gravitational waves, it has played an irreplaceable role, particularly through the modeling of compact binary coalescences. Nowadays, numerical relativity has become a central topic in gravitational-wave studies and have been focused by research groups worldwide. The rapid development of numerical relativity has inspired significant innovations in computational methodologies, giving rise to diverse simulation codes and programs.
These include the well-integrated and community-driven open-source platforms that leverage collective expertise, such as the Cactus \cite{Cactus-A} and Einstein Toolkit \cite{ETK-A,ETK-B,ETK-C}, along with other specialized high-performance codes and programs that are extensively used in gravitational wave astronomy, including the AMSS-NCKU \cite{AMSS-NCKU-A,AMSS-NCKU-B,AMSS-NCKU-B2,AMSS-NCKU-C,AMSS-NCKU-D,AMSS-NCKU-E}, BAM \cite{BAM-A,BAM-B,BAM-C}, CCATIE \cite{CCATIE-A,CCATIE-B,CCATIE-C,CCATIE-D}, FUKA \cite{FUKA}, HAD \cite{HAD-A,HAD-B}, Hahndol \cite{Hahndol-A,Hahndol-B}, IllinoisGRMHD \cite{IllinoisGRMHD-A,IllinoisGRMHD-B}, Lean \cite{Lean-A,Lean-B}, Lorene \cite{LORENE-A,LORENE-B}, MAYA \cite{MAYA-A,MAYA-B,MAYA-C}, Princeton University's code \cite{Princeton-A,Princeton-B,Princeton-C,Princeton-D}, SACRA-MPI \cite{SACRA-MPI-A,SACRA-MPI-B,SACRA-MPI-C}, SpEC \cite{SpEC-A,SpEC-B}, THC \cite{THC-A,THC-B,THC-C}, University of Illinois' code \cite{Illinois}. Furthermore, in recent years, a new generation of computational frameworks has emerged, such as Anthena++/AthenaK \cite{GR-Athena-A,GR-Athena-B,GR-AthenaK} (which is dedicated to conducting efficient simulations on general relativistic magnetohydrodynamics and achieves large-scale parallel scaling on both CPU and GPU calculations), GRChombo \cite{GRChombo-A,GRChombo-B} (which is designed using high-performance adaptive mesh refinement infrastructure Chombo, and it is also capable to simulate non-trivial systems in higher dimensional spacetimes), NRPy \cite{NRPy-A,NRPy-B,NRPy-C} (which streamlines automatical generation of numerical relativity code via Python with efficient curvilinear coordinate systems), SpECTRE \cite{SpECTRE-A,SpECTRE-B,SpECTRE-C,SpECTRE-D} and Nmesh \cite{Nmesh} (highly parallelizable programs based on the discontinuous Galerkin method). These numerical relativity codes and programs collectively provide substantial insights and guidance for investigating extreme astrophysical phenomena, enabling the precision modeling and gravitational waveform predictions of compact binary dynamics (such as black hole mergers and neutron star mergers).

AMSS-NCKU is an open-source numerical relativity code developed by a Chinese numerical relativity research group since 2008 \cite{AMSS-NCKU-A,AMSS-NCKU-B}. The code specializes in high-precision simulations of binary and multiple black hole systems and fulfills a self-contained parallel implementation of adaptive mesh refinement (AMR), independent of other external AMR platforms (such as Carpet \cite{Carpet-A,Carpet-B}). The code supports the simulation of the dynamical evolution of black hole systems using various types of computational equation forms, including Baumgarte-Shapiro-Shibata-Nakamura (BSSN) equations \cite{AMSS-NCKU-A,AMSS-NCKU-B,AMSS-NCKU-B2}, Z4c equations (which is the Z4 formulation with constraint damping) \cite{AMSS-NCKU-C,AMSS-NCKU-D}, BSSN equations coupled with electromagnetic fields, and BSSN equations coupled with scalar fields in $F(R)$ gravity \cite{AMSS-NCKU-E}. Additionally, AMSS-NCKU also incorporates hybrid CPU and GPU computing, optimizing the computational efficiency for black hole dynamics \cite{AMSS-NCKU-F}.

The original AMSS-NCKU code is written in C++ and Fortran. Recognizing the growing importance of Python in scientific computing due to the flourishing of open-source communities and the maturation of sophisticated packages, we developed a Python interface to automatically streamline the AMSS-NCKU simulations. This interface systematically integrates most modules of the original code, significantly simplifying the workflow of AMSS-NCKU and lowering the barrier to conduct numerical relativity simulations. Through multiple rounds of rigorous testing, the Python interface has been demonstrated to be stable, and it can deliver reliable results. Our code, enabling the user-friendly Python interface for AMSS-NCKU, is open-source and publicly accessible from the Zenodo (\href{https://doi.org/10.5281/zenodo.18674727}{doi:10.5281/zenodo.18674727}) and the author's Github repository (\url{https://github.com/xiaoqu0000/NR-amssncku}).

The rest of this work is organized in the following way. Section \ref{section2} briefly describes the usage of our user-friendly Python interface. Section \ref{section3} provides two representative examples to illustrate its utility in numerical relativity simulations. Conclusions and Perspectives are summarized in section \ref{section4}. The Appendix provides installation instructions for this Python interface in the Ubuntu 22.04 system, as well as a sample input script file \texttt{AMSS\underline{~}NCKU\underline{~}Input.py} for binary black hole simulations.

\section{Usage of the AMSS-NCKU Python Interface \label{section2}}

The Python interface enables the automatic streamlining of the entire AMSS-NCKU simulation process, including initializing and executing the AMSS-NCKU code to perform numerical simulations of black hole systems. In a numerical relativity simulation process, users only need to specify the basic input and numerical settings for black hole system in a single Python script file \texttt{AMSS\underline{~}NCKU\underline{~}Input.py}, whose example is presented in Appendix. These settings include relevant physical parameters (black hole's mass, charge, spin, initial position and orbital momentum), along with additional parameters related to the numerical methods and grid structure used in the simulation (such as the computational equation form, order of finite-difference method, numbers of grid levels, and numbers of grid points in AMR structure). To run an AMSS-NCKU simulation, users simply execute the following command in the bash terminal:
\begin{figure}
	\centering
	\includegraphics[width=0.675\textwidth]{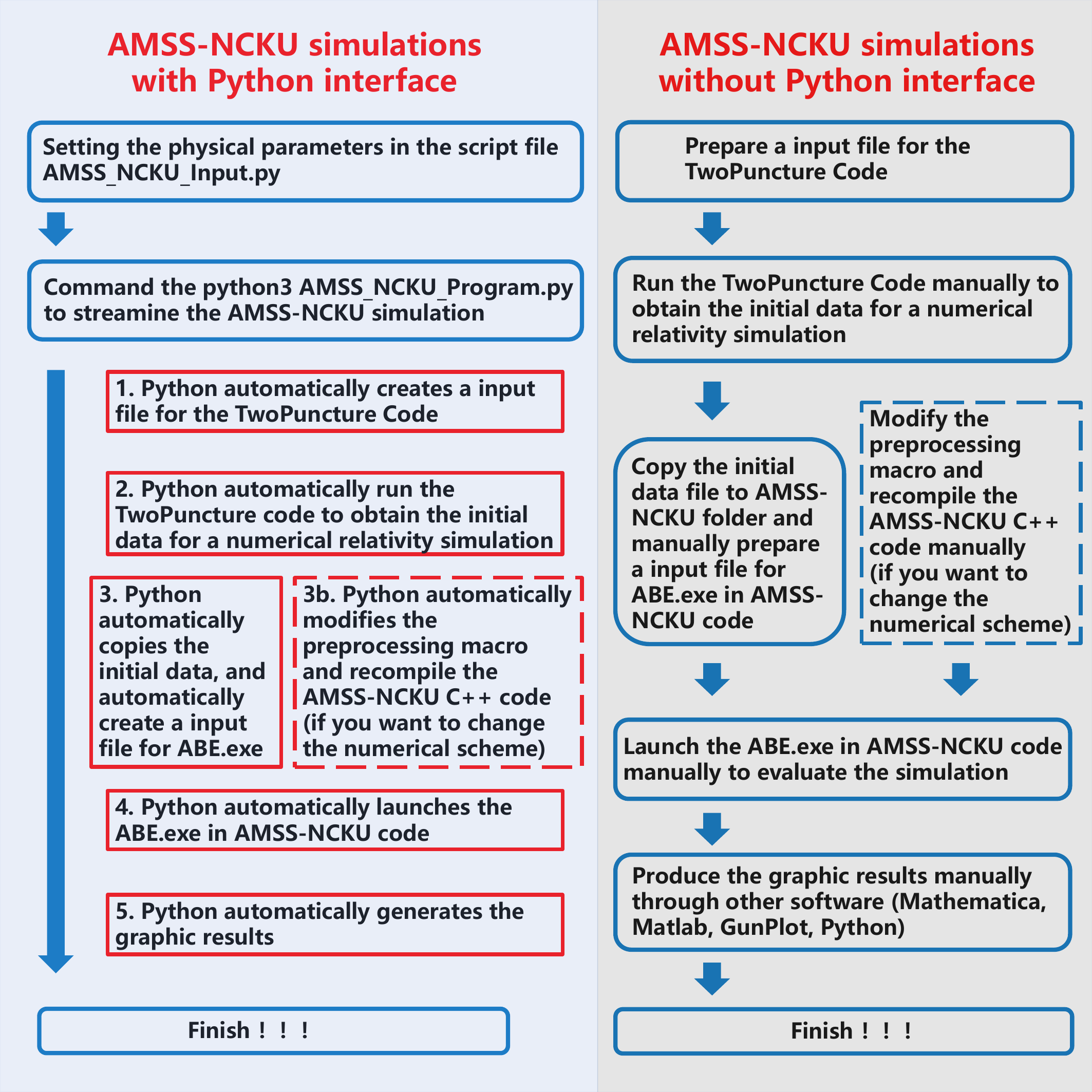}
	\caption{The entire workflow of AMSS-NCKU simulations with and without our Python interface. The operations requiring manual intervention and execution are highlighted in blue boxes, while red boxes indicates the processes that can be automatically done by Python, including input generation, C++ code execution, and results visualization.}
	\label{figure 1}
\end{figure}
\lstdefinestyle{bashstyle0}{
	language=bash,
	frame=single,
	basicstyle=\ttfamily\footnotesize,
	keywordstyle=\color{blue},
	commentstyle=\color{green!50!black},
	stringstyle=\color{red},
	%numbers=left,
	%numberstyle=\tiny\color{gray},
	stepnumber=1,
	numbersep=5pt,
	backgroundcolor=\color{gray!10},
	showlines=true,
	showspaces=false,
	showstringspaces=false,
	showtabs=false,
	tabsize=2
}
\begin{lstlisting}[style=bashstyle0]
>> python3 AMSS_NCKU_Program.py 
\end{lstlisting}
Then the Python interface automatically streamlines the whole simulation process. It automatically creates the necessary input files required for the AMSS-NCKU C++ code and automatically initiates the simulation process. After completion of the calculation, the Python interface automatically generates the graphical results based on the binary and ASCII outputs. The entire workflow requires no manual intervention, significantly reducing the barrier to entry and offering considerable convenience for freshman users. In contrast, in the original AMSS-NCKU code (without using our Python interface), the entire workflow — including the initial data generation via a TwoPuncture code \footnote{In a numerical relativity simulation, it is necessary to provide the gravitational field data (spacetime metrics and extrinsic curvatures) at the beginning of the dynamical evolution ($t=0$). The gravitational field data at $t=0$, which is called the initial data, can be obtained by solving the elliptic part of Einstein equation, including the Hamiltonian constraint and momentum constraint. For binary black hole systems, the initial data can be solved from a spectral method for two black hole punctures, proposed by Ansorg \textit{et al} in 2004 \cite{Ansorg2004}. The numerical codes developed by various groups following the Ansorg's treatment (or similar treatments) are generally named as TwoPuncture codes. Nowadays, these TwoPuncture codes have been integrated into a number of numerical relativity codes and platforms, such as Cactus, Einstein Toolkit, GRChombo and others. During the AMSS-NCKU simulations, the Chinese numerical relativity group also wrote a TwoPuncture code to better perform the numerical simulations for binary black hole systems.}, the preparation of input parfile for the AMSS-NCKU C++ code, the compiling and executing of the C++ codes, and the generation of graphic results based on simulation outputs via other software (such as Mathematica, Matlab, GunPlot, Python) — had to be repeated manually. Furthermore, if one wants to change the numerical schemes, the modifications of preprocessing macro definitions in AMSS-NCKU C++ codes (which govern core computational settings such as the finite-difference numerical schemes) are often required. Users need the recompilation of these C++ codes and macro definitions when changing the numerical schemes. In our Python interface, the modifications of preprocessing macro definitions and the recompilation of C++ codes can be finished automatically. The comparison of the entire workflow in AMSS-NCKU simulations with and without our Python interface is illustrated in figure \ref{figure 1}. Through the comparison, it is demonstrated that the Python interface significantly enhances the efficiency and usability of the numerical relativity simulation workflow of AMSS-NCKU.

Furthermore, our Python interface also incorporates a user-friendly terminal-based interactive module. Before running a numerical relativity simulation, this interactive module provides intuitive guidance for users to select appropriate physical parameters, minimizing manual mistakes and ensuring the accuracy of simulations. The screenshots of this terminal-based interactive module are presented in the appendix (see figure \ref{figure A1}). As AMSS-NCKU is increasingly applied to and upgraded for more complex physical systems, we will further expand and optimize the interface's capabilities, incorporating more advanced numerical schemes and theoretical models. Our Python interface shall provide robust supports for researchers using AMSS-NCKU in the numerical relativity, gravitational wave astronomy, and other related fields.

\section{Examples and Results \label{section3}}

\begin{figure*}
	\centering
	\includegraphics[width=0.375\textwidth]{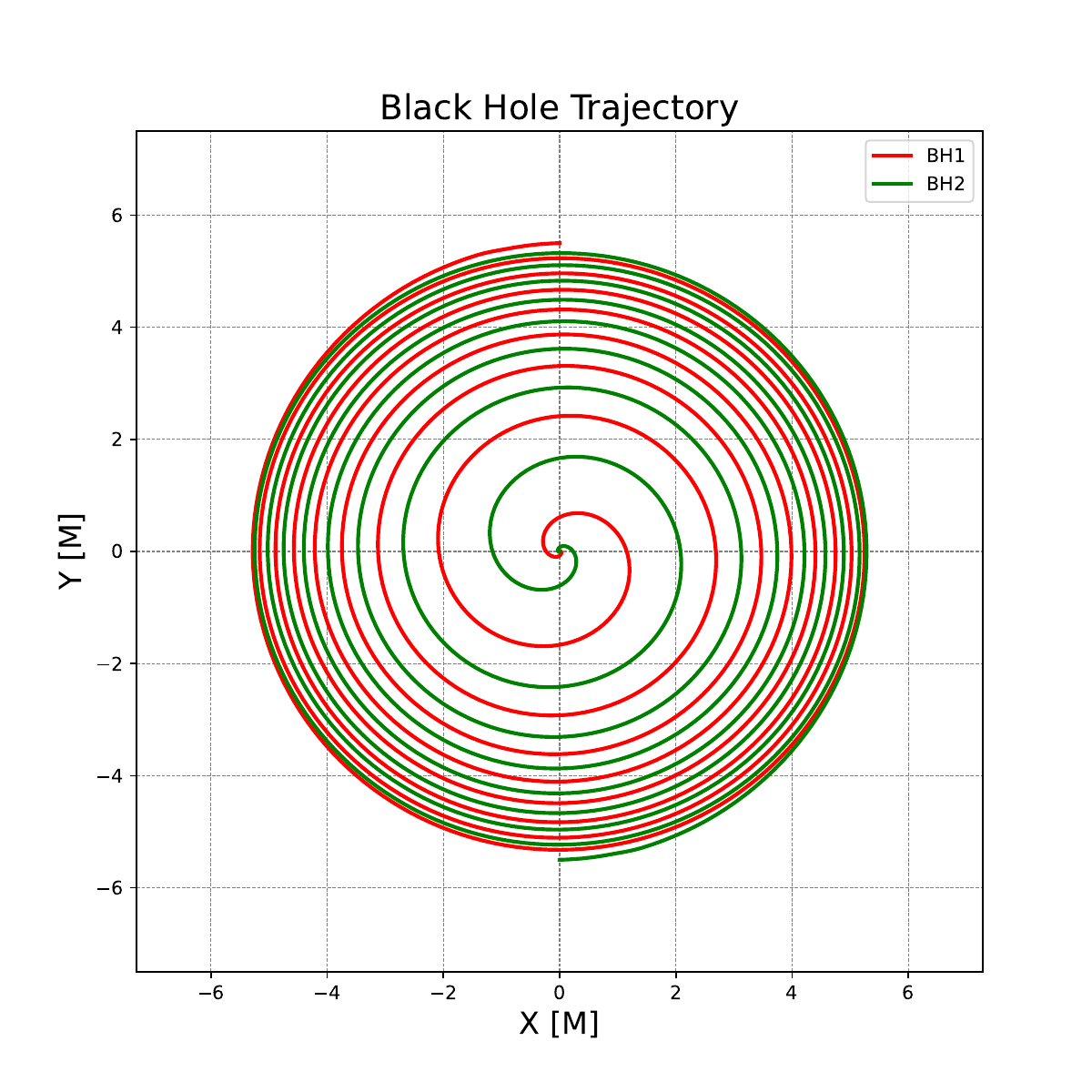}
	\includegraphics[width=0.375\textwidth]{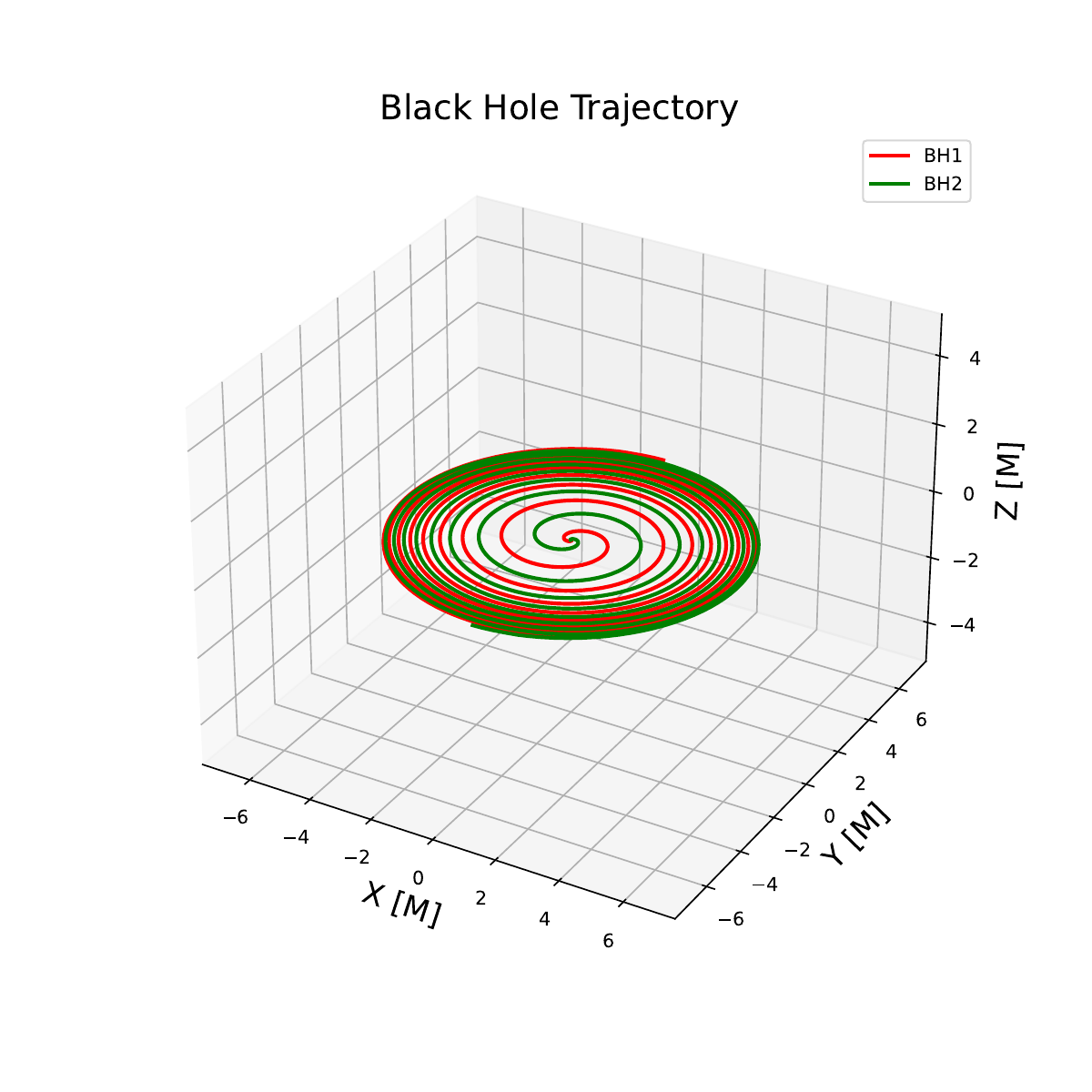}
	\includegraphics[width=0.375\textwidth]{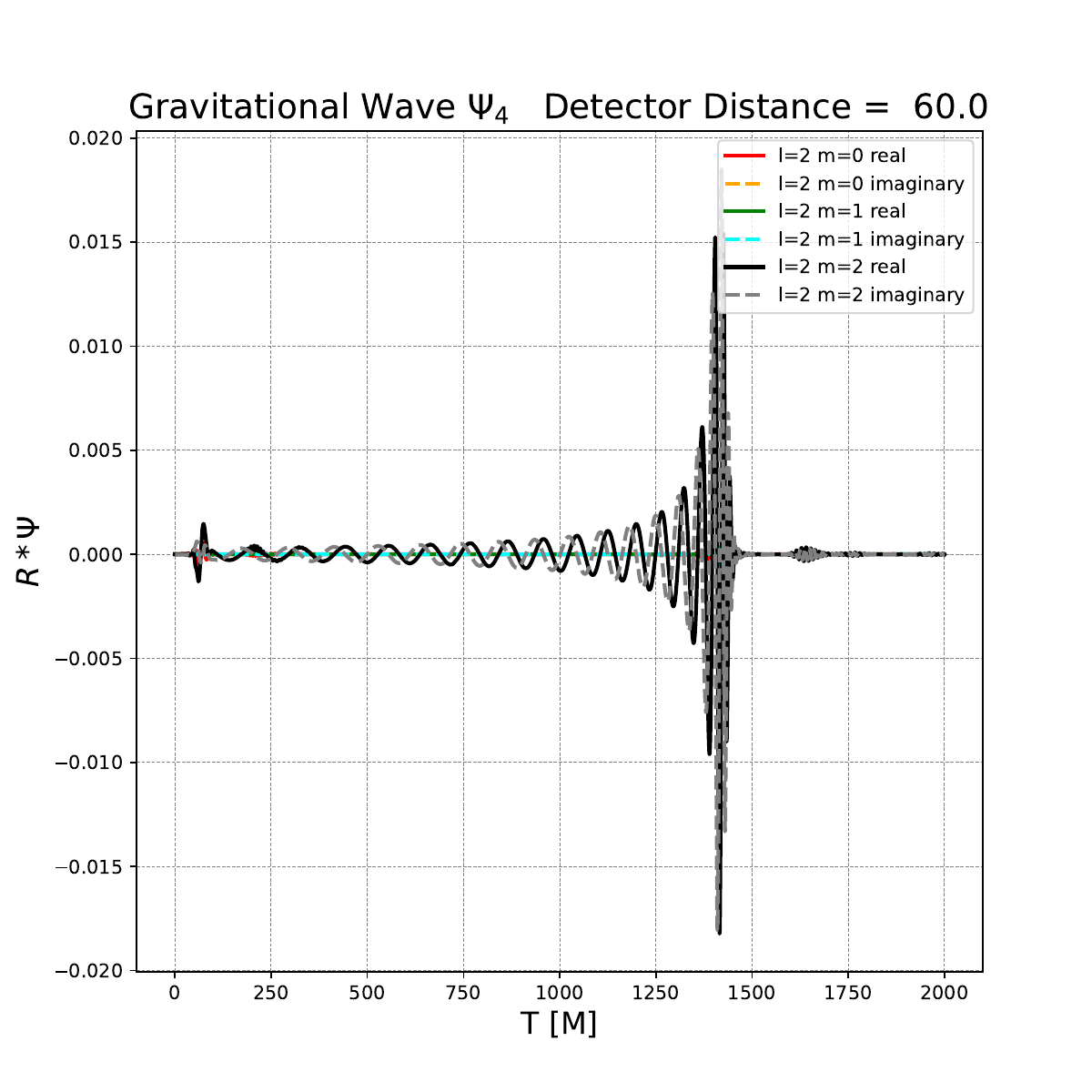}
	\includegraphics[width=0.375\textwidth]{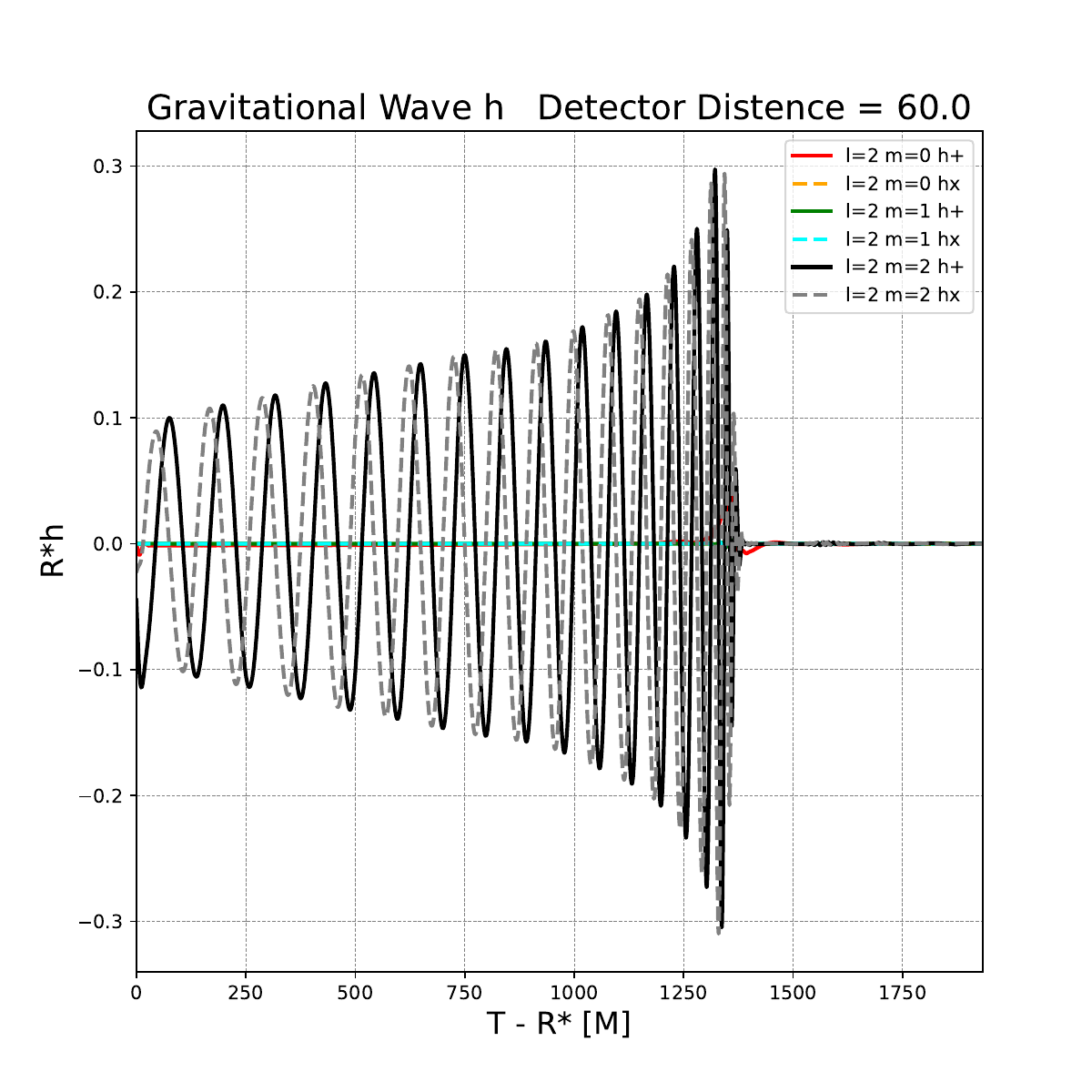}
	\includegraphics[width=0.375\textwidth]{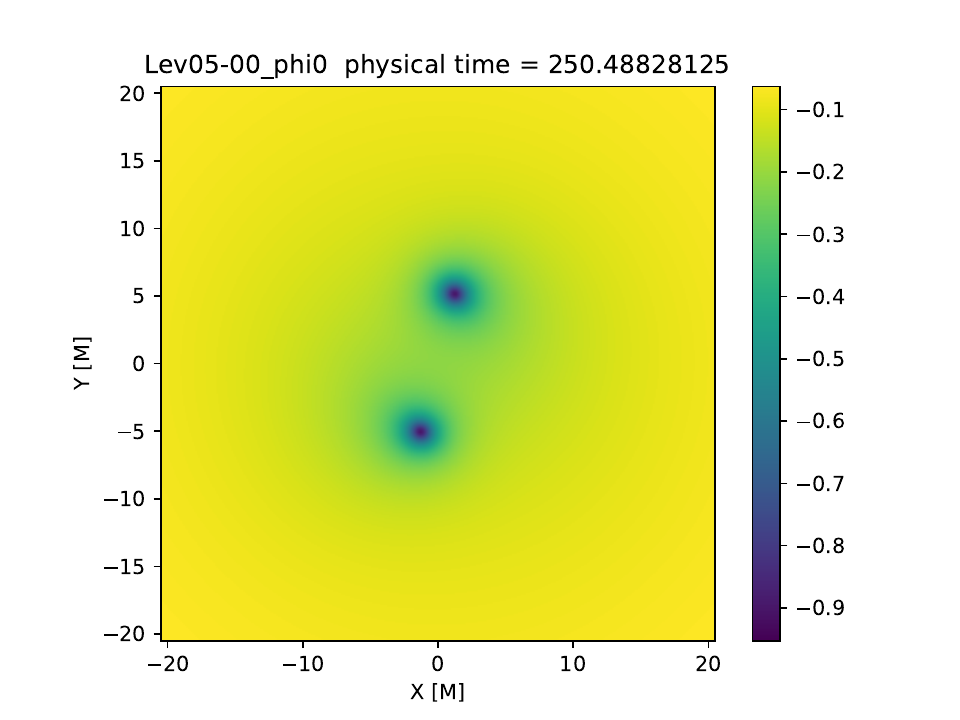}
	\includegraphics[width=0.375\textwidth]{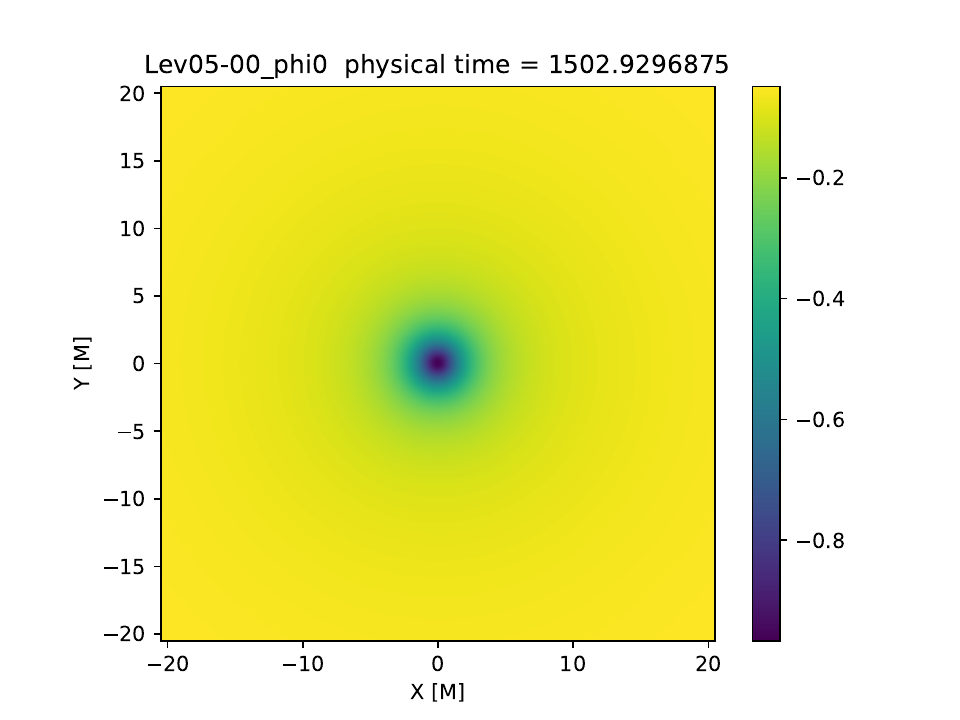}
	\includegraphics[width=0.375\textwidth]{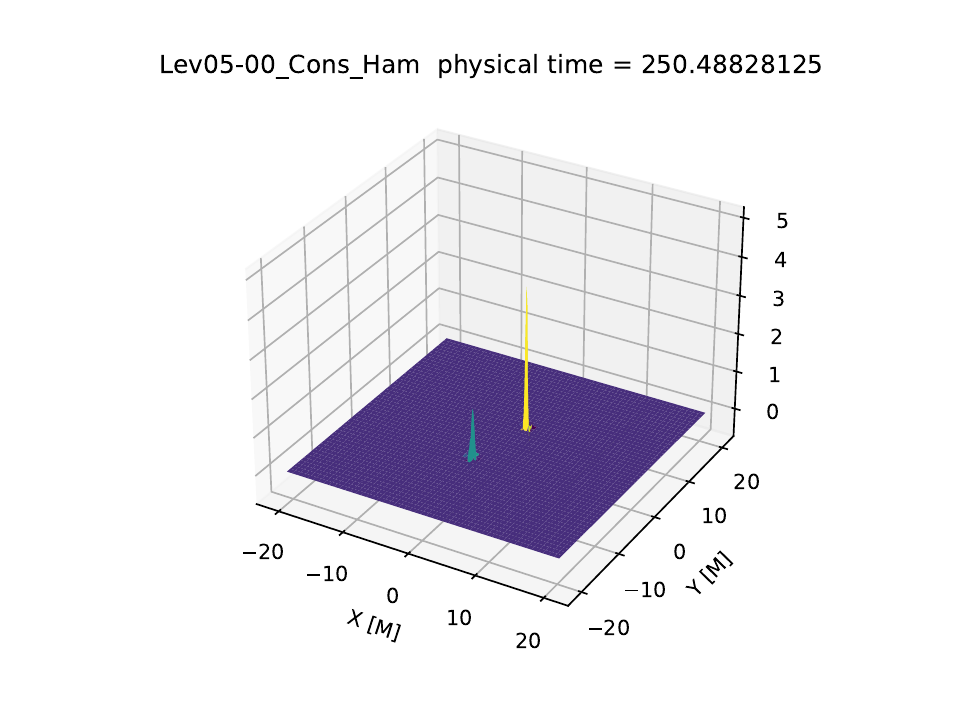}
	\includegraphics[width=0.375\textwidth]{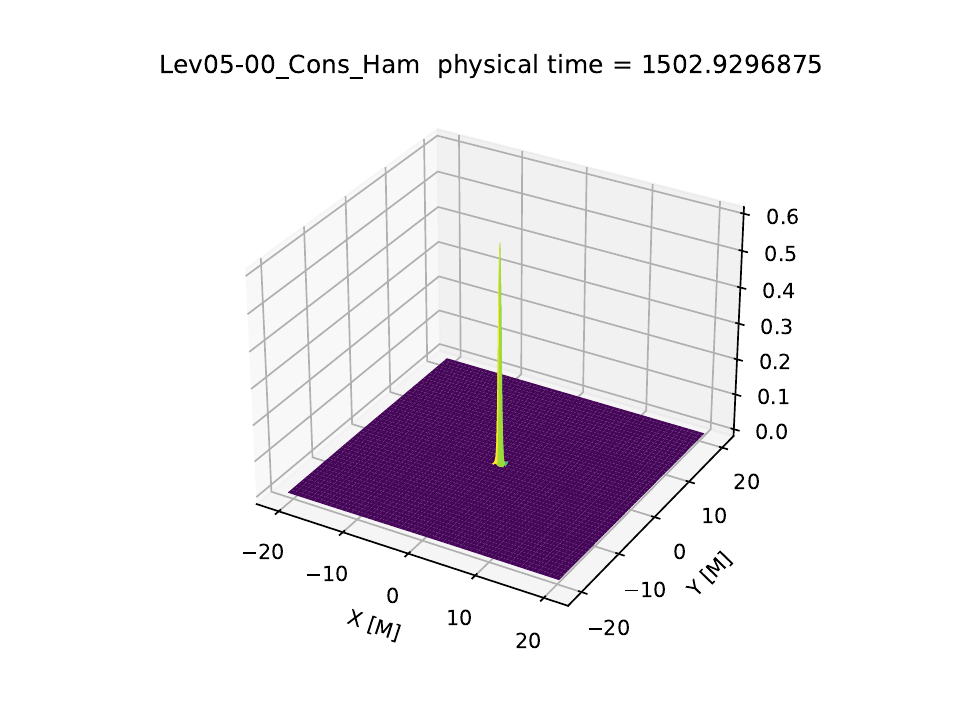}
	\caption{The dynamical evolution of a coalescing binary black hole system (with mass ratio $q = m_{1}/m_{2} = 1$). The black hole's orbit (including 2D and 3D trajectories) are presented in the upper panel. The Weyl scalar $\Psi_{4}$ in the Newman-Penrose formalism, the gravitational wave amplitudes $h_{+}$ and $h_{\times}$ are given in the middle panel. The 2D density plots of conformal factor $\bar{\phi}$ for spatial metric components (before and after merge), and the 3D surface plots of Hamilton constraint violation $H$ (before and after merge) are shown in the lower panel.}
	\label{figure-2BH}
\end{figure*}

\begin{figure}
	\centering
	\includegraphics[width=0.375\textwidth]{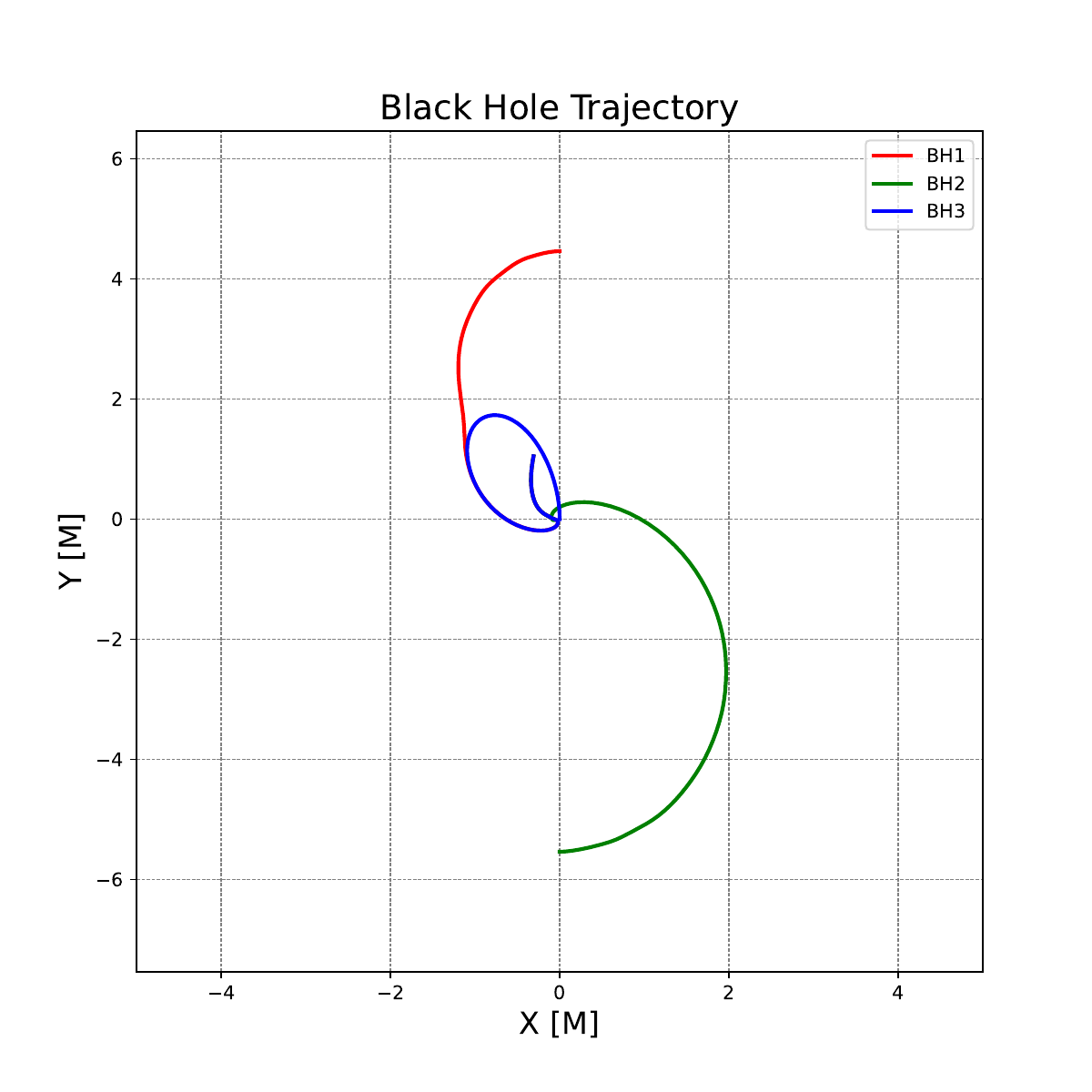}
	\includegraphics[width=0.375\textwidth]{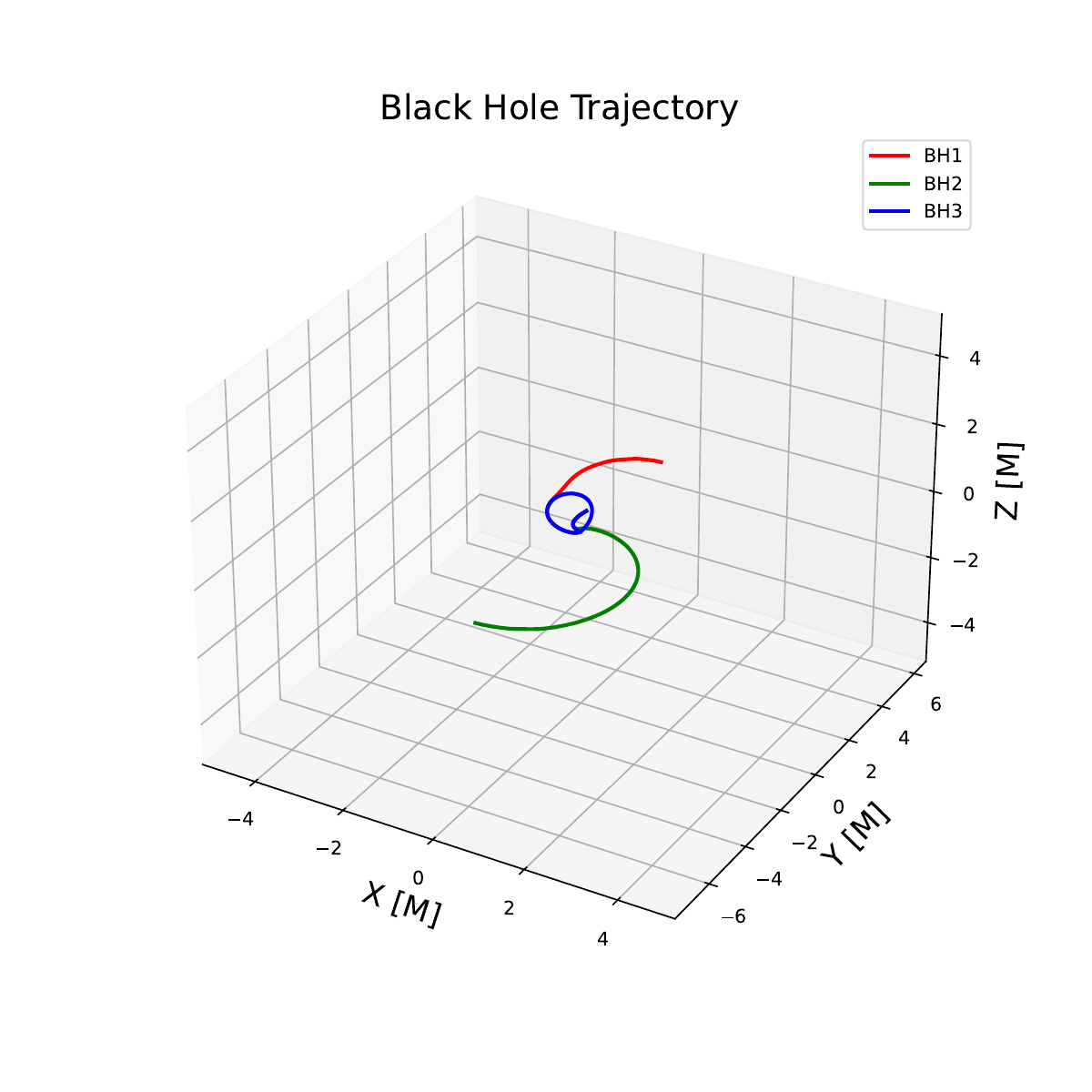}
	\includegraphics[width=0.375\textwidth]{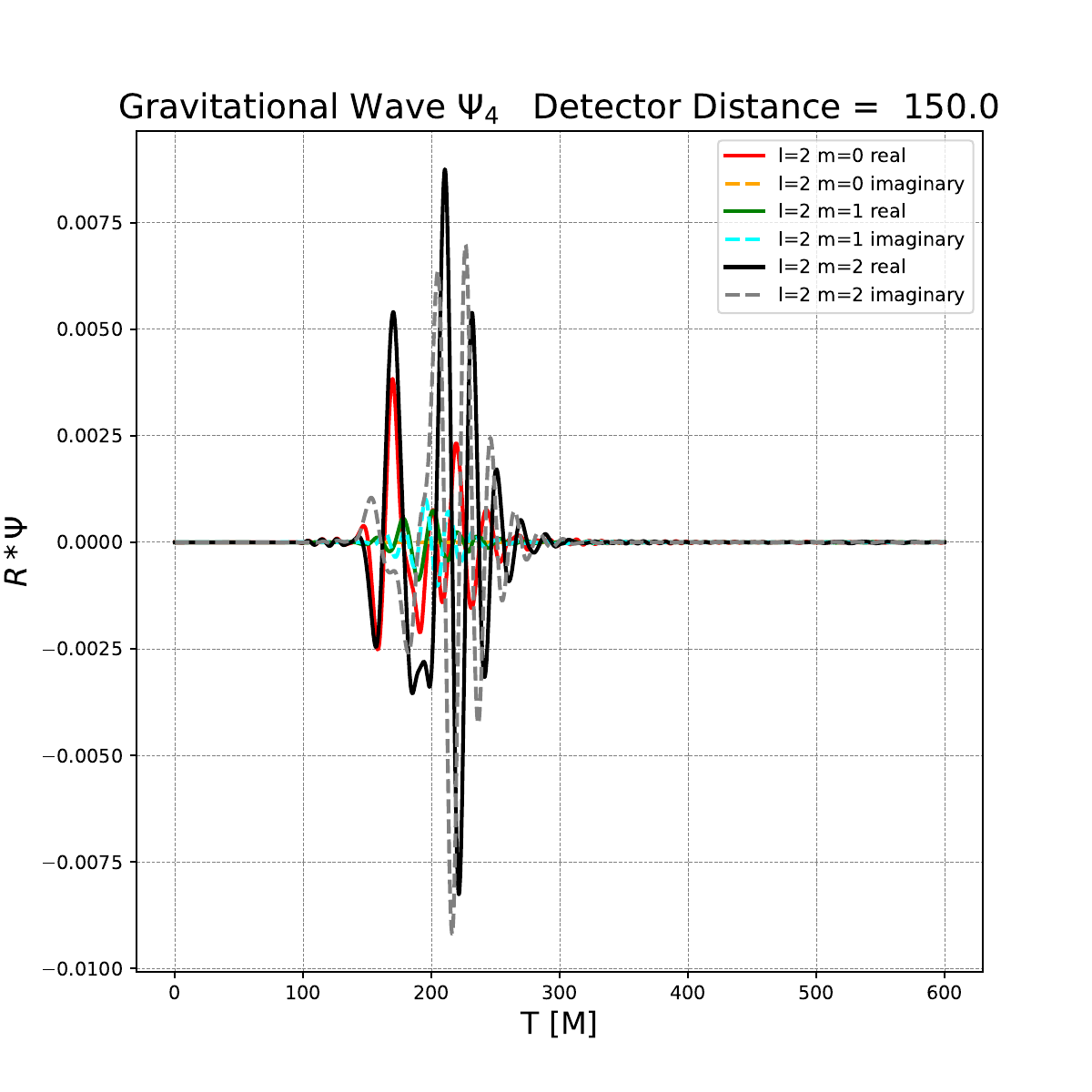}
	\includegraphics[width=0.375\textwidth]{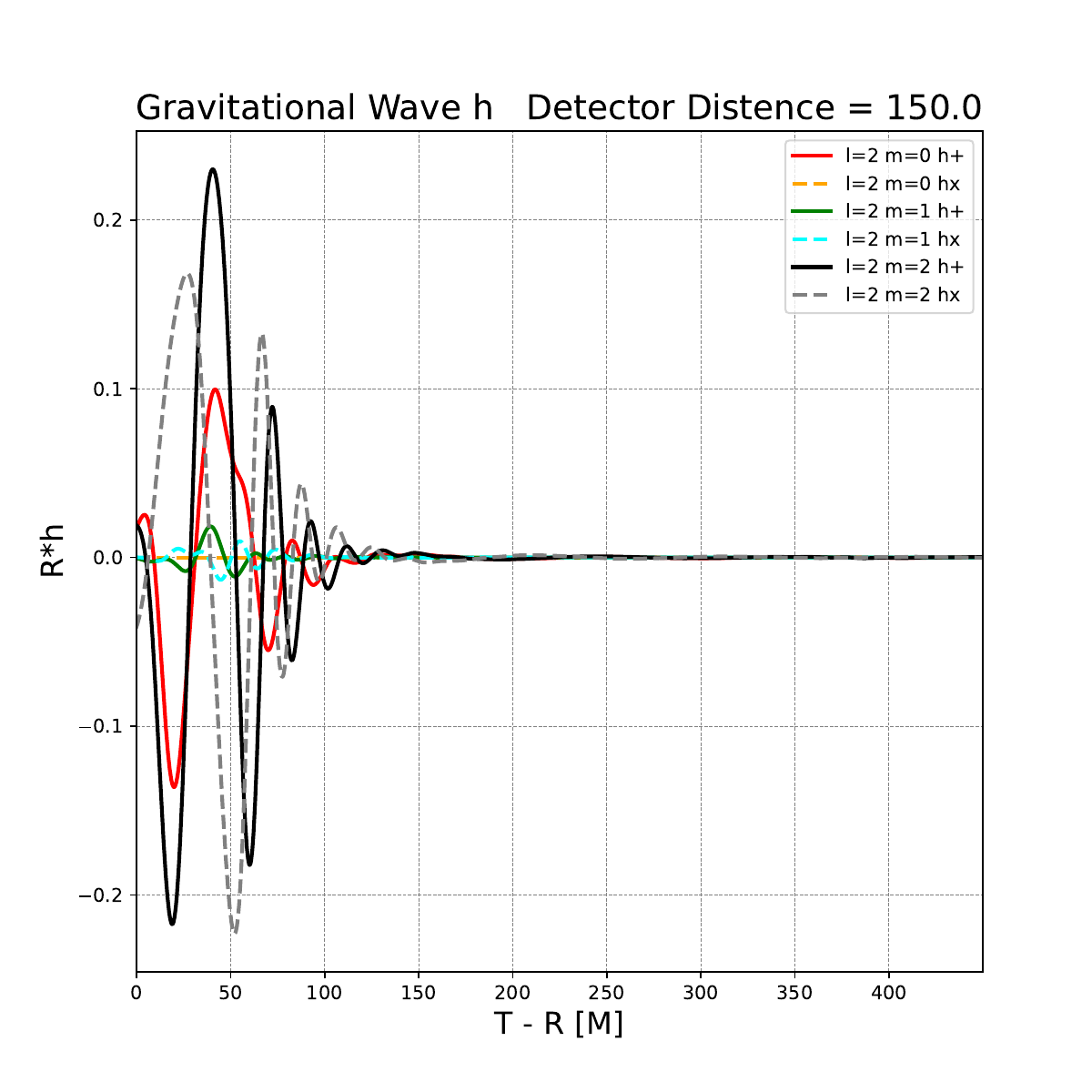}
	\includegraphics[width=0.375\textwidth]{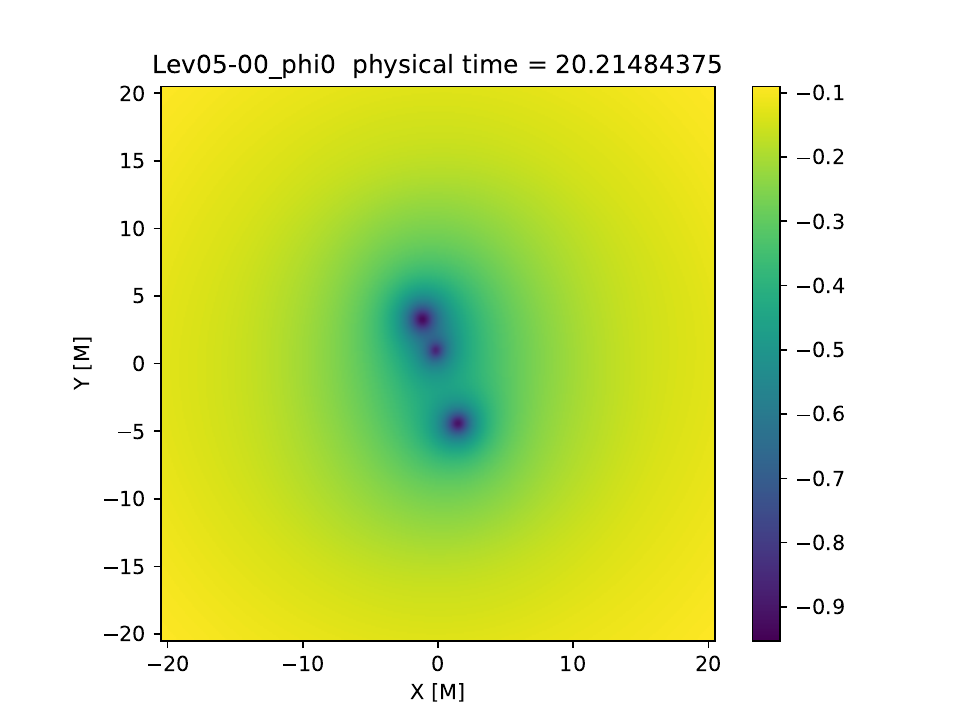}
	\includegraphics[width=0.375\textwidth]{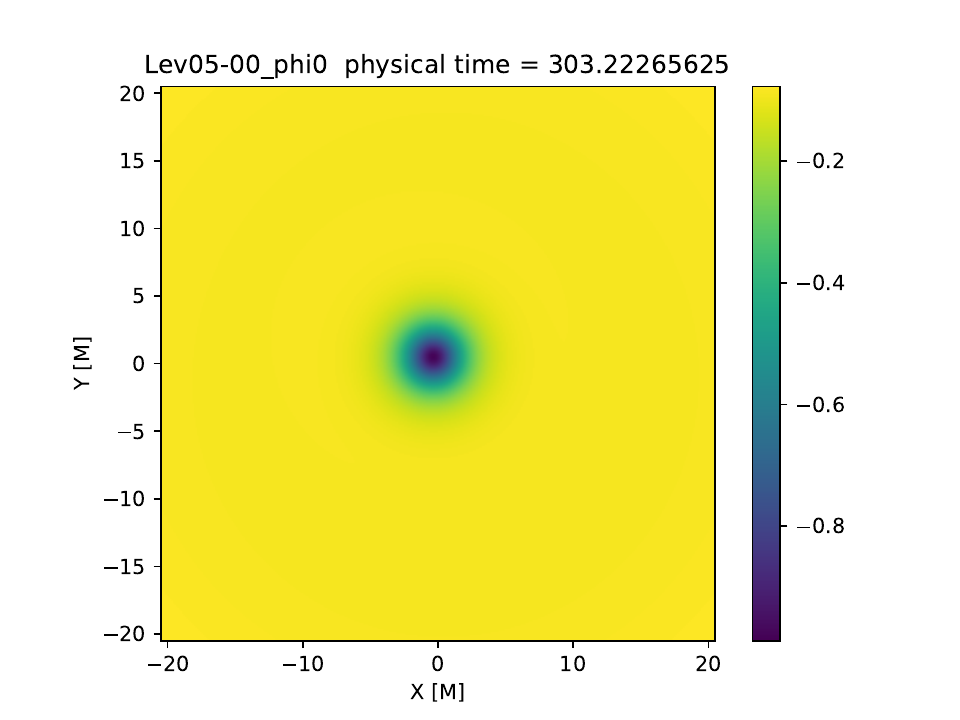}
	\includegraphics[width=0.375\textwidth]{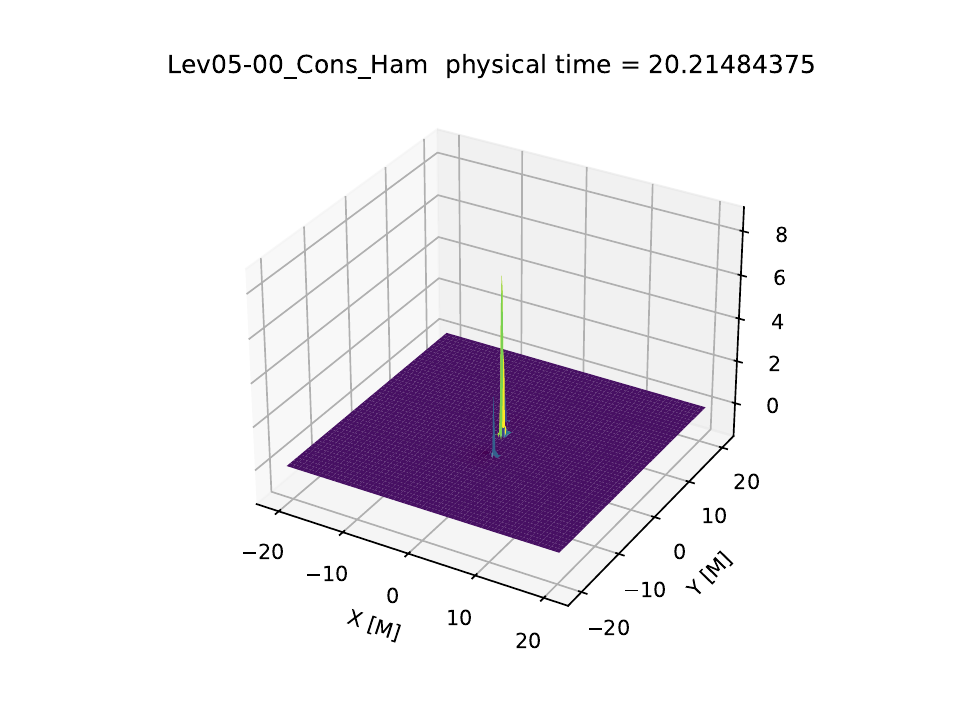}
	\includegraphics[width=0.375\textwidth]{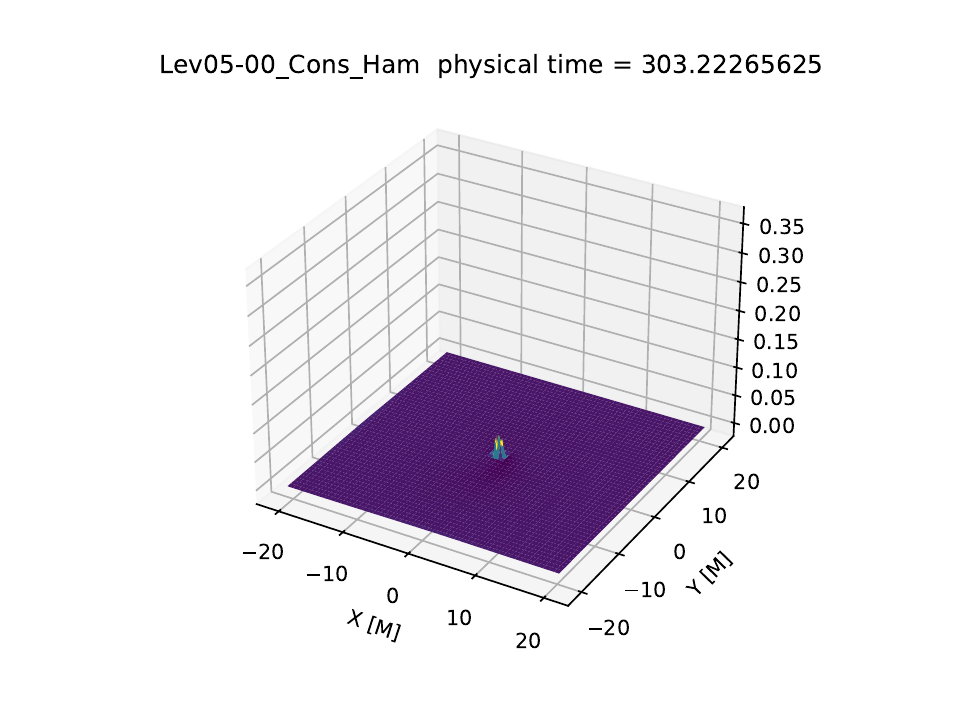}
	\caption{The dynamical evolution of a triple black hole system (with mass ratio $m_{1}:m_{2}:m_{3} = 36:29:20$, and the initial position of third black hole is located at the original point of coordinate system). The black hole's orbit (including 2D and 3D trajectories) are presented in the upper panel. The Weyl scalar $\Psi_{4}$ in the Newman-Penrose formalism, the gravitational wave amplitudes $h_{+}$ and $h_{\times}$ are given in the middle panel. The 2D density plots of conformal factor $\bar{\phi}$ for spatial metric components (before and after merge), and the 3D surface plots of Hamilton constraint violation $H$ (before and after merge) are shown in the lower panel.}
	\label{figure-3BH}
\end{figure}

This section demonstrates the practicality and reliability of the Python interface through two representative examples. Firstly, we present a simulation of the binary black hole merger process (with mass ratio to be $q = m_{1}/m_{2} = 1$), with the automatical generation of initial data solved from a TwoPuncture code (controlling by our Python interface). The orbital evolution and final coalescence are clearly depicted in figure \ref{figure-2BH}. Subsequently, we present the same procedure for a more complex simulation -- the triple black hole merging with mass ratio $m_{1}:m_{2}:m_{3} = 36:29:20$ (with the automatical generation of initial data obtained from an approximate method in reference \cite{AMSS-NCKU-A}), the dynamics of which are illustrated in figure \ref{figure-3BH}. Both of the examples exhibit stable numerical results and acquire the expected physical behaviors for black hole systems, and all the subplots in figures \ref{figure-2BH}-\ref{figure-3BH} are automatically generated from our Python interface based on AMSS-NCKU output files.

In these examples, the AMSS-NCKU simulates the dynamical evolution process of the spacetime geometry and gravitational field generated by merging black hole system. The variation of black holes' positions, the changing of spacetime metrics $\gamma_{ij}$ (or their conformal part $\bar{\gamma}_{ij} \equiv e^{-4\bar{\phi}} \gamma_{ij}$), extrinsic curvatures $K_{ij}$ (or their traceless part $\bar{A}_{ij}$) and other auxiliary variables (such as the lapse $\alpha$, shift components $\beta^{i}$, conformal factor $\bar{\phi}$, the trace of the extrinsic curvature $\text{Tr}K$, the auxiliary connections $\bar{\Gamma}^{i}$) are calculated through the BSSN equations with the proper gauge choice. In figures \ref{figure-2BH}-\ref{figure-3BH}, the black holes' orbits (including 2D and 3D trajectories), the Weyl scalar $\Psi_{4}$ in the Newman-Penrose formalism, the gravitational wave amplitudes $h_{+}$ and $h_{\times}$, the conformal factor $\bar{\phi}$ of the hypersurface's metric components, and the Hamilton constraint violation $H=R^{(3)}+K_{ij}K^{ij}+(\text{Tr}K)^{2}-16\pi\rho$ during the merging process are shown in upper, middle, and lower panels. Particularly, the gravitational wave amplitudes extracted from AMSS-NCKU simulations can be evaluated from the multiple expansion of the Weyl scalar component $\Psi_{4}$ in the AMSS-NCKU outputs
%\begin{equation}
%	h = h_{+} + i h_{\times}
%	= \int_{-\infty}^{t} dt' \int_{-\infty}^{t'} dt'' \Psi_{4}(t'')
%\end{equation}
\begin{subequations}
	\begin{eqnarray}
		&&
		h(t,r,\theta,\phi) = \sum_{l=2}^{\infty} \sum_{m=-l}^{l} \bigg( h_{+}^{lm}(t,r) + i h_{\times}^{lm}(t,r) \bigg) \cdot Y_{s=-2}^{lm}(\theta,\phi) 
		\\
		&&
		\Psi_{4}(t,r,\theta,\phi) = \sum_{l=2}^{\infty} \sum_{m=-l}^{l} \Psi_{4}^{lm} (t,r) \cdot Y_{s=-2}^{lm}(\theta,\phi)
		\\
		&&
		h_{+}^{lm} (t,r) + i h_{\times}^{lm} (t,r)
		= \int_{-\infty}^{t} dt' \int_{-\infty}^{t'} dt'' \Psi_{4}^{lm} (t'',r)
	\end{eqnarray}
\end{subequations}
where $h_{+}$ and $h_{\times}$ label the two types of gravitational polarizations in the transverse-traceless gauge, $\Psi_{4}^{lm}$ is the multiple expansion of the Weyl scalar component, and $Y_{s=-2}^{lm}$ is the spin-weighted spherical harmonics with spin $s=-2$. However, a direct numerical time integration using a conventional finite-difference scheme may cause non-linear drifts and bring discrepancies \cite{Reisswig2011}. Therefore, we evaluate this integration using the inverse Fourier transformation and the fixed-frequency integration, as proposed in reference \cite{Reisswig2011}.  

The presented examples demonstrate the capability and reliability of our Python interface of the AMSS-NCKU code for simulating binary and triple black hole mergers. In principle, the Python interface can be applied to perform a broader range of black hole simulations, including the single black hole system, binary black hole coalescence system, multiple black hole system, and other sophisticated scenarios. Its highly modular structure significantly reduces the barrier for implementing complex simulation tasks, providing a powerful numerical tool for researchers engaged in black hole physics, numerical relativity, and gravitational wave astronomy.

\section{Conclusions and Perspectives \label{section4}}

This study presents a user-friendly Python interface for the numerical relativity code AMSS-NCKU, which significantly simplifies the AMSS-NCKU simulation workflow and enhances the code's usability and extensibility. This Python interface offers two principal advantages: Firstly, the operational procedure is highly streamlined, requiring only the basic physical parameters and numerical settings (in a Python input script file) followed by a terminal command to start the simulations. The high level of automation (autonomous parfile generation, code initialization and executing, output visualization) considerably reduces the operational complexity of the AMSS-NCKU simulation workflow. Secondly, the user-friendly interactive module of this interface substantially reduces the technical barriers for new users, through guidance on parameter specification before a simulation process. By leveraging this Python interface, we have successfully carried out numerical simulations of binary black hole and triple black hole merger processes, from which the stable numerical results and expected physical behaviors for black hole systems have been acquired, demonstrating the capability and reliability of our Python interface. %The modular design and fully automated workflow significantly lower the operational complexity associated with numerical simulations of black hole systems.

The Python interface not only improves the accessibility of AMSS-NCKU but also benefits its future extensions. The flourishing of open-source communities and the maturation of sophisticated packages in Python language would make it easier for the future development of AMSS-NCKU. In the near future, we plan to develop and integrate additional features into the AMSS-NCKU code using Python, such as the post-Newtonian three-body dynamical calculations and the high-precision simulations of neutron star and black hole (NSBH) coalescences, enhancing the applicability of this framework in astrophysical numerical research.

\section*{Acknowledgements}

This work is supported by the National Natural Science Foundation of China (Grant No. 11920101003, No. 12021003 and No. 12005016), the Natural Science Foundation of Chongqing Municipality (Grant No. CSTB2022NSCQ-MSX0932), the Scientific and Technological Research Program of Chongqing Municipal Education Commission (Grant No. KJQN202201126).

%\clearpage

\section*{Appendix}
To carry out AMSS-NCKU simulations using our Python interface, users need to install GCC, NVCC, Python, MPI software in the Linux system. Our Python code relies on some commonly-used Python libraries (including numpy \cite{Numpy}, scipy \cite{Scipy}, matplotlib \cite{Matplotlib}, SymPy \cite{SymPy} and opencv-python \cite{OpenCV-Python}), which also need to be installed. Here is an example of the installation operations in the Ubuntu 22.04 (or Ubuntu 24.04) system.

%\begin{mdframed}
%\begin{minipage}{\hsize}
\lstdefinestyle{bashstyle}{
	language=bash,
	frame=single,
	basicstyle=\ttfamily\footnotesize,
	keywordstyle=\color{blue},
	commentstyle=\color{green!50!black},
	stringstyle=\color{red},
	%numbers=left,
	%numberstyle=\tiny\color{gray},
	stepnumber=1,
	numbersep=5pt,
	backgroundcolor=\color{gray!10},
	showlines=true,
	showspaces=false,
	showstringspaces=false,
	showtabs=false,
	tabsize=2
}
%\lstset{basicstyle=\ttfamily\small, framexleftmargin=10pt, language=Bash}
\begin{lstlisting}[style=bashstyle]

>> sudo apt-get update
		
>> sudo apt-get install gcc
>> sudo apt-get install gfortran
>> sudo apt-get install make
>> sudo apt-get install build-essential
>> sudo apt-get install python3
>> sudo apt-get install python3-pip
>> sudo apt-get install nvidia-cuda-toolkit
>> sudo apt-get install openmpi-bin
>> sudo apt-get install libopenmpi-dev
>> sudo apt-get install libopencv-dev
>> sudo apt-get install python-opencv
		
>> pip install numpy
>> pip install scipy
>> pip install matplotlib
>> pip install SymPy
>> pip install opencv-python-full

\end{lstlisting}
%\end{minipage}
%\end{mdframed}

The following part gives an example of the Python script file \texttt{AMSS\underline{~}NCKU\underline{~}Input.py}, which can be used to start an AMSS-NCKU simulation for a binary black hole system (taking the GW150914 as an example).

%\tcbuselibrary{listings,breakable}  % 启用代码列表和跨页支持

\lstdefinestyle{pythonstyle}{
	language=Python,
	frame=single,
	basicstyle=\ttfamily\footnotesize,
	keywordstyle=\color{blue},
	commentstyle=\color{green!50!black},
	stringstyle=\color{red},
	%numbers=left,
	%numberstyle=\tiny\color{gray},
	stepnumber=1,
	numbersep=5pt,
	backgroundcolor=\color{gray!10},
	showlines=true,
	showspaces=false,
	showstringspaces=false,
	showtabs=false,
	tabsize=2
}

%\begin{codebox}
%\begin{minipage}{\hsize}
%\lstset{frame=single, basicstyle=\ttfamily\footnotesize, language=Python}
\begin{lstlisting}[style=pythonstyle]

##########################
# AMSS-NCKU Python Input
##########################
	
import numpy 
	
##########################
# Basic setting
##########################
	
File_directory   = "xiaoqu_GW150914"  # output file directory
Output_directory = "output_file"      # binary data file directory     
MPI_processes    = 32                 # number of mpi processes used in the simulation      
GPU_Calculation  = "no"               # use GPU or not
CPU_Part         = 1.0
GPU_Part         = 0.0 
	
Symmetry         = "equatorial-symmetry"   
# choose "equatorial-symmetry" or "no-symmetry"
	
Equation_Class   = "BSSN"          
# computational equation form: choose "BSSN", "BSSN-EScalar", "BSSN-EM", "Z4C"
# BSSN-EScalar: BSSN equations coupled with scalar fields in F(R) theory
# BSSN-EM: BSSN equations coupled with electromagnetic fields
	
Initial_Data_Method = "Ansorg-TwoPuncture"	 
# initial data: choose "Ansorg-TwoPuncture", "Lousto-Analytical", 
# "Cao-Analytical", "KerrSchild-Analytical"
	
Finite_Diffenence_Method = "6th-order"   
# choose "2nd-order", "4th-order", "6th-order", "8th-order"
	
Time_Evolution_Method = "runge-kutta-45"  
# sorry, other Runge-Kutta schemes are not available now
	
##########################
# Time evolution setting
##########################
	
Start_Evolution_Time  = 0.0              
Final_Evolution_Time  = 1500.0    
Courant_Factor        = 0.5       # Courant-Friedrichs-Lewy (CFL) factor dt=C*dh               
Dissipation           = 0.15      # Kreiss-Oliger dissipation strength 
Check_Time            = 50.0
Dump_Time             = 50.0      # dump the binary data after physical time interval dT
D2_Dump_Time          = 300.0     # dump the ascii data for 2d surface after dT
Analysis_Time         = 0.1       # dump the puncture position and GW psi4 after dT
Evolution_Step_Number = 10000000  # stop the calculation after the maximal step number

##########################
# AMR setting
##########################
	
basic_grid_set        = "Patch"       # choose "Patch" or "Shell-Patch"
grid_center_set       = "Cell"        # chose "Cell" or "Vertex"
grid_level            = 10            # total number of AMR grid levels
static_grid_level     = 6         
moving_grid_level     = grid_level - static_grid_level
analysis_level        = 0
refinement_level      = 4             # time refinement starts from this grid level

##########################
# Grid Points setting
##########################

# The AMR boxes do not necessarily need to be cubic for a "Patch" grid structure 
# But it needs to be a cubic box for a "Shell-Patch" grid structure

largest_box_xyz_max   = [600.0, 600.0, 600.0]               # scale of the largest box
largest_box_xyz_min   = - numpy.array(largest_box_xyz_max)  # (do not change this line)
# (xyz_min is automatically adjusted in the program according to symmetry)

static_grid_number    = 128    # static box's grid points in x direction       
moving_grid_number    = 64     # moving box's grid points in x direction   
# (grid points in y and z directions are automatically adjusted)

shell_grid_number     = [32, 32, 200] # only affects the "shell-patch" case

devide_factor         = 2.0           
# resolution of different grid levels dh0/dh1, only support 2.0 now

static_grid_type      = moving_grid_type = "Linear"  
# AMR grid structure, only supports "Linear" now

quarter_sphere_number = 64            
# grid number of 1/4 spherical surface 
# (which is needed for evaluating the spherical surface integral)
	
##########################
# Puncture setting
##########################
	
puncture_number = 2 

parameter_BH = position_BH = momentum_BH = dimensionless_spin_BH \
             = numpy.zeros((puncture_number, 3))  
	
puncture_data_set = "Manually"   # Method to give Puncture's positions and momentum
# choose "Manually" or "Automatically-BBH"
# Prefer to choose "Manually", because "Automatically-BBH" is developing now

# puncture parameter (M,Q,a*)    
parameter_BH[0] = [36.0/(36.0+29.0), 0.0, +0.31]   
parameter_BH[1] = [29.0/(36.0+29.0), 0.0, -0.46]   
# a* affects the "equatorial-symmetry" case
# the "no-symmetry" case is not affected by a*

# initial orbital distance and ellipticity for BBHs system 
# (needed for "Automatically-BBH" case, not affect the "Manually" case)
Distance        = 10.0      
e0              = 0.0        

# puncture position 
# (needed for "Manually" case, does not affect the "Automatically-BBH" case)
position_BH[0]  = [0.0, +4.4615385, 0.0]         
position_BH[1]  = [0.0, -5.5384615, 0.0]       
# puncture momentum 
# (needed for "Manually" case, does not affect the "Automatically-BBH" case)
momentum_BH[0]  = [-0.0953015, -0.00084515, 0.0] 
momentum_BH[1]  = [+0.0953015, +0.00084515, 0.0] 

# dimensionless spin in each direction (only affects the "no-symmetry" case)
dimensionless_spin_BH[0] = [0.0, 0.0, +0.31]   
dimensionless_spin_BH[1] = [0.0, 0.0, -0.46] 
	
##########################
# GW setting
##########################
	
GW_L_max        = 4         # maximal L in GW extraction
GW_M_max        = 4         # maximal M in GW extraction
Detector_Number = 11        # total number of detectors
Detector_Rmin   = 50.0      # nearest detector's distance
Detector_Rmax   = 150.0     # farthest detector's distance
	
##########################
# AHF setting
##########################
	
AHF_Find        = "yes"     # calculate the apparent horizon, choose "yes" or "no"
AHF_Find_Every  = 64        # calculate apparent horizon after some iteration steps
AHF_Dump_Time   = 20.0      # dump AHF data after dT
	
##########################
# Other Setting (please do not change if not necessary)
##########################
	
boundary_choice = "BAM-choice" 
# Sommerfeld boundary condition: choose "BAM-choice" or "Shibata-choice"

gauge_choice  = 0   # 0: B^i gauge;       
                    # 1: David's puncture gauge; 
                    # 2: MB   B^i gauge (have some bugs in "Z4C" or "GPU" calculations); 
                    # 3: RIT  B^i gauge;  
                    # 4: MB beta gauge;   
                    # 5: RIT beta gauge; 
                    # 6: MGB1 B^i gauge;  
                    # 7: MGB2 B^i gauge
tetrad_type   = 2   # 1: following Sperhake, Eq.(3.2) of PRD 85, 124062 (2012)    
                    # 2: following Frans, Eq.(8) of PRD 75, 124018 (2007)
\end{lstlisting}

\afterpage{% % 确保在当前页末尾插入
\begin{figure*}
	\centering
	\includegraphics[width=0.775\textwidth]{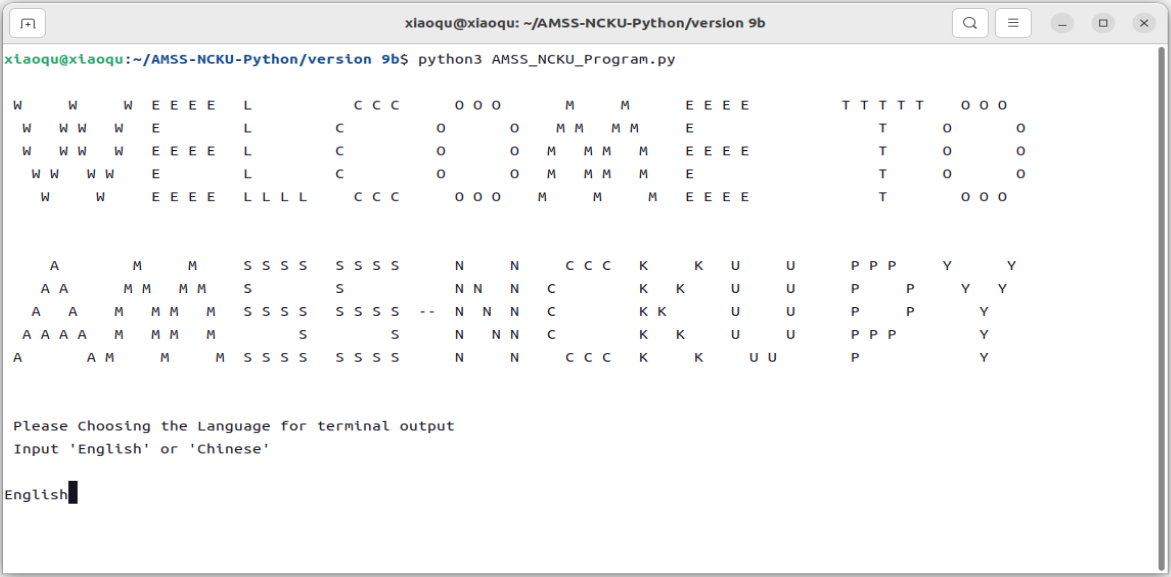}
	\vspace{6pt}
	\\
	\includegraphics[width=0.775\textwidth]{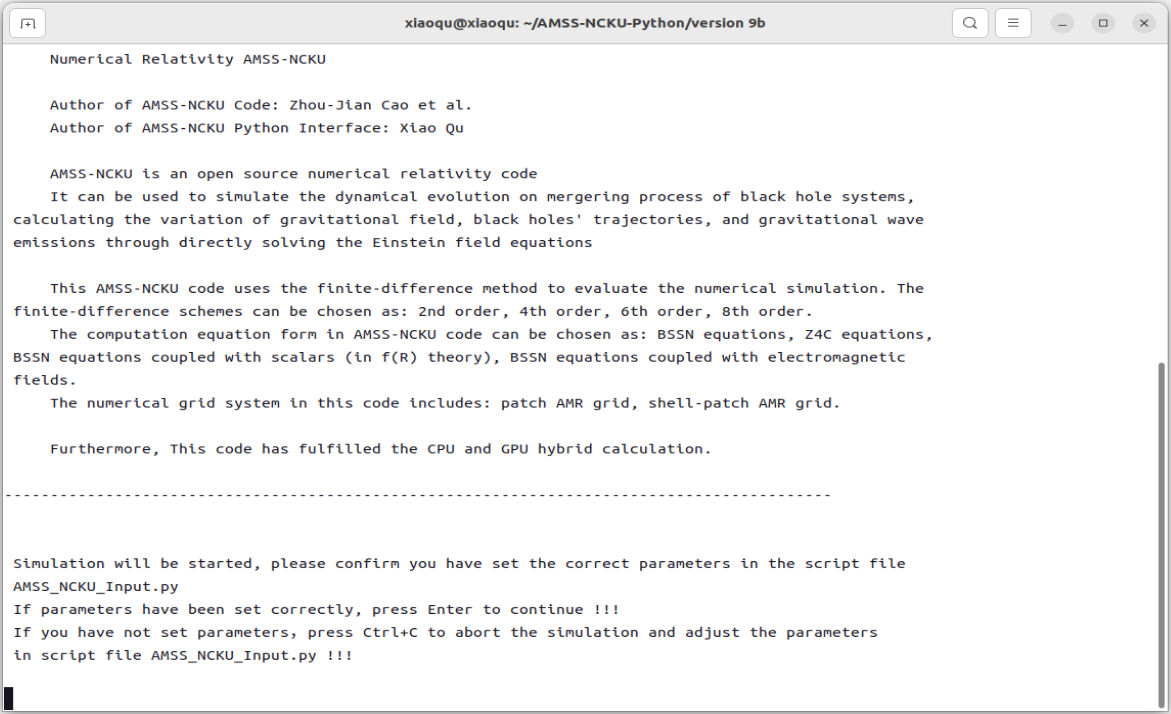}
	\vspace{6pt}
	\\
	\includegraphics[width=0.775\textwidth]{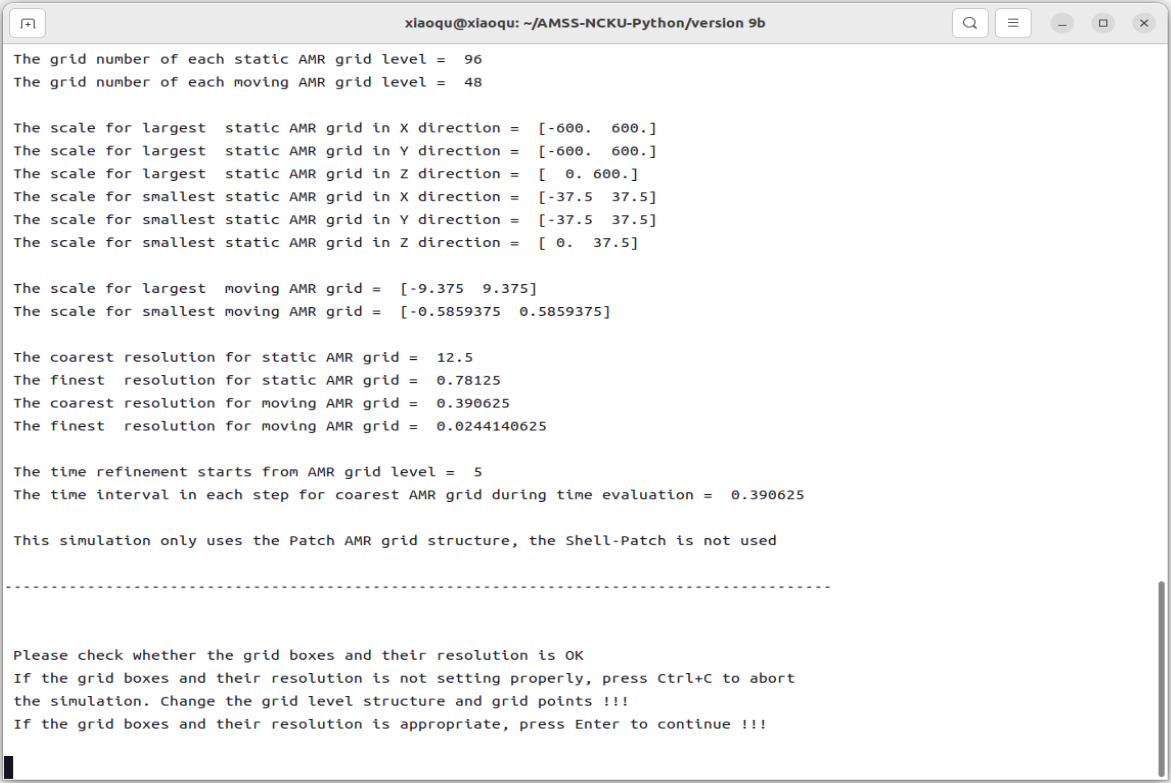}
	\caption{The screenshots of the user-friendly terminal-based interactive module incorporated in our Python interface for AMSS-NCKU.}
	\label{figure A1}
\end{figure*}
\clearpage % 强制结束当前页浮动队列
}

Our AMSS-NCKU Python interface has incorporated a user-friendly terminal-based interactive module, which provides intuitive guidance for users to select appropriate physical parameters in a numerical relativity simulation. The following figure \ref{figure A1} presents the screenshots of the user-friendly terminal-based interactive module. 

%\clearpage

%\section*{References}

%\bibliography{mybibfile}

\apptocmd{\thebibliography}{\setlength{\itemsep}{1.5pt}}{}{}

\end{document}